\documentclass{emulateapj}
\usepackage{xspace}
\usepackage{amsmath}
\usepackage{hyperref}
\usepackage{framed} 
\usepackage{txfonts}
\usepackage{tikz}

\def\HII{{\rm HII}}
\def\qi{Q_\HII}
\def\Msun{M_\odot}

\def\dim#1{\mbox{\,#1}}
\def\hide#1{}

\begin{document}

\title{Warm Dark Matter and Cosmic Reionization}

\author{Pablo Villanueva-Domingo\altaffilmark{1}}
\author{Nickolay Y.\ Gnedin\altaffilmark{2,3,4}}
\author{Olga Mena\altaffilmark{1}} 
\altaffiltext{1}{Instituto de F\'isica Corpuscular (IFIC), CSIC-Universitat de Val\`encia,\\
Apartado de Correos 22085,  E-46071, Spain}
\altaffiltext{2}{Particle Astrophysics Center, Fermi National Accelerator Laboratory, Batavia, IL 60510, USA; gnedin@fnal.gov}
\altaffiltext{3}{Kavli Institute for Cosmological Physics, The University of Chicago, Chicago, IL 60637 USA;}
\altaffiltext{4}{Department of Astronomy \& Astrophysics, The
  University of Chicago, Chicago, IL 60637 USA} 

\begin{abstract}
In models with dark matter made of particles with keV masses, such as a sterile neutrino, small-scale density perturbations are suppressed, delaying the period at which the lowest mass galaxies are formed and therefore shifting the reionization processes to later epochs. In this study, focusing on Warm Dark Matter (WDM) with masses close to its present lower bound, i.e. around the  $3$~keV region, we derive constraints from galaxy luminosy functions, the ionization history and the Gunn-Peterson effect. We show that even if star formation efficiency in the simulations is adjusted to match the observed UV galaxy luminosity functions in both CDM and WDM models, the full distribution of Gunn-Peterson optical depth retains the strong signature of delayed reionization in the WDM model. However, until the star formation and stellar feedback model used in modern galaxy formation simulations is constrained better, any conclusions on the nature of dark matter derived from reionization observables remain model-dependent.
\end{abstract}


\maketitle

\section{Introduction}
\label{sec:intro}
The current canonical cosmological model, the so-called $\Lambda$-Cold Dark Matter ($\Lambda$CDM) scenario, assumes that dark matter is made of a totally cold weakly interacting particles, representing $\sim26\%$ of the universe mass-energy density. This very simple framework provides an accurate description of both large scale structure observations \citep{Alam:2016hwk} and the temperature and polarization anisotropies of the Cosmic Microwave 
Background (CMB) measured with unprecedented precision by the Planck satellite \citep{Ade:2015xua}.
However, the precise microphysics and, ultimately, the nature of dark matter remains undiscovered \citep{Bertone:2004pz,Bergstrom:2012fi}. Particle physics models cover a very wide range for possible  masses, interactions, and free-streaming scales for the dark matter particles. The last property is probably the most interesting one for cosmological and astrophysical observations. Free-streaming allows the dark matter particles to propagate out of density perturbations, suppressing the growth of structure at scales smaller than a characteristic scale. For practical purposes, such a scale is totally irrelevant for CDM particles. However, there are a number of well-motivated hypothesized particles (such as sterile neutrinos) for which its free-streaming nature is very relevant and can not be neglected when computing our universe's structure formation history. 
 
Besides that particle physics provides excellent \emph{theoretically}-motivated dark matter candidates with non-negligible free-streaming scales,
there also exist several \emph{observational} indications that are sometimes interpreted as the so-called "small-scale crisis" of the 
 $\Lambda$CDM model \citep[see][for the recent review]{Bullock:2017xww}, which further motivate the search for
 alternatives to the standard CDM scenario: abundances and kinematics of gravitationally
 bound structures predicted in purely collisionless, not accounting for any baryonic physics numerical simulations of the $\Lambda$CDM paradigm are not consistent with several observational measures of analogous quantities in the Local Group and other nearby galaxies \citep{Klypin:1999uc,Moore:1999nt,BoylanKolchin:2011dk}.
 Plenty of work has been devoted to trying to resolve these existing controversies at galactic and
sub-galactic scales \citep{Wang:2016rio,Lovell:2016nkp,Sawala:2012cn,Sawala:2015cdf,Fattahi:2016nld,Polisensky:2013ppa,Vogelsberger:2012ku,Schewtschenko:2015rno,Lovell:2011rd,Lovell:2013ola,Lovell:2015psz,Nakama:2017ohe}. Among the possible solutions, a Warm Dark Matter (WDM) candidate with a $\sim$ keV mass will have a non-negligible free-streaming scale, providing one of several potential explanations. Furthermore, claims of the detection of a monochromatic line at $3.56$~keV in X-ray data toward the Andromeda and Perseus Cluster were reported in Refs.~\cite{Bulbul:2014sua,Boyarsky:2014jta}. Such a line has been identified as the radiative decay of a WDM keV sterile neutrino into a photon plus an active neutrino state ($\nu_s \to \gamma \nu$). The Ly-$\alpha$ power spectra from distant quasars has been shown to provide a robust tool where to test and extract the most constraining bounds on the mass of a thermal WDM relic \citep{Irsic:2017ixq,Yeche:2017upn,Viel:2005qj,Seljak:2006qw,Viel:2006kd,Viel:2007mv,Boyarsky:2008xj,Viel:2013apy,Baur:2015jsy}, albeit it is subject to potential uncertainties in modeling of dynamics and thermodynamics of the  
Inter Galactic Medium (IGM) at low and intermediate redshifts. 

In this work we focus on the constraints on the WDM mass around the $3$~keV region (where  current constraints lie \citep{Irsic:2017ixq,Yeche:2017upn}) from the reionization period  \citep{Barkana:2001gr,Yoshida:2003rm,Somerville:2003sh,Yue:2012na,Pacucci:2013jfa,Mesinger:2013nua,Schultz:2014eia,Dayal:2015vca,Lapi:2015zea,Bose:2016hlz,Bose:2016irl,Corasaniti:2016epp,Menci:2016eui,Lopez-Honorez:2017csg}, when the ultraviolet photons emitted by the first forming galaxies ionized the neutral IGM. With its free-streaming nature, WDM particles wash out small-scale overdensities, delaying the formation of the lowest mass galaxies (which are among the main sources of ionizing photons), and, consequently, shifting the reionization processes to later epochs.

We make use of extensive hydrodynamical simulations, which model the complex physics involved in the cosmic reionization processes. Therefore, we self-consistently account for a wide range of physical effects such as star formation and stellar feedback, non-equilibrium ionization of hydrogen and helium, etc \citep{Gnedin:2014uta}, several of them highly degenerate with the WDM free-streaming nature. We then compare the results arising from our hydrodynamical simulations to a set of reionization-related observables. We consider measurements of the UV galaxy luminosity functions \citep{Bouwens:2014fua}, the ionization fraction of the universe \citep{Fan:2005es}, and distributions of the Gunn-Peterson optical depth \citep{Becker:2014oga}. 

All in all, our calculations show that, while there is little hope in distinguishing among several possible WDM scenarios and the canonical $\Lambda$CDM case using current astronomical measurements of the observables listed above, future observations with the James Webb Space Telescope \citep[JWST, ][]{Gardner:2006ky} may reach the required sensitivity to differentiate WDM from CDM in reionization observables, if systematic uncertainties due to currently poorly known reionization ingredients, such as stellar feedback, can be reduced and kept under control by, for example, future progress in constraining galaxy formation models.
 
\section{WDM cosmologies}
WDM is characterized by having non-negligible velocities at high redshifts, suppressing the growth of structures on scales below a free-streaming length, which depends on the mass of the WDM particle $m_X$ and it is typically of hundred kiloparsecs for keV WDM candidates. The free-streaming scale can be computed as the distance over which such a particle can travel until the time of matter radiation equality $t_{\rm eq}$ \citep{Kolb:1990vq, Schneider:2011yu}, while more accurate predictions require numerical simulations \citep{Bode:2000gq}. In contrast to the standard WIMP scenario, WDM particles are still relativistic at the decoupling epoch, but non-relativistic at $t_{\rm eq}$, when substantial growth of perturbations becomes possible. Hereafter we consider that all the dark matter is warm and that it was thermally decoupled in the early universe. Predictions for non-thermal relics as non-resonantly produced sterile neutrinos can be easily related to our results \citep[c.f.][]{Viel:2005qj}.

\begin{figure}
\includegraphics[width=\hsize]{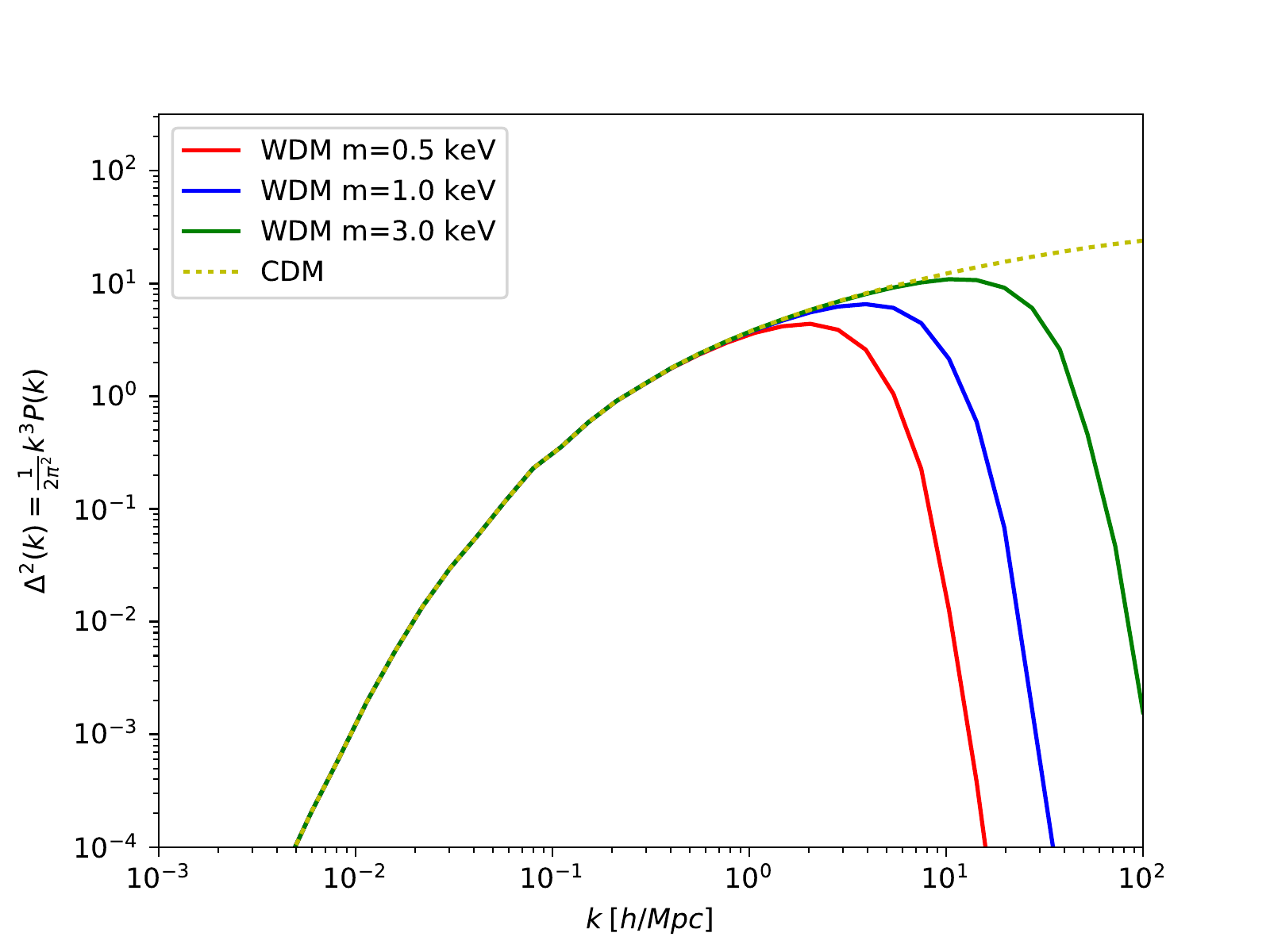}
\caption{Dimensionless linear power spectrum $\Delta^2(k) = \frac{1}{2 \pi^2} k^3 P(k) $ for several values of the WDM particle mass $m_X$, together with the $\Lambda$CDM case.\label{fig:delta}}
\end{figure}

The WDM free-streaming nature leads to a CDM-like power spectrum with a cutoff, which is around a scale $10^{10}$ ($10^{8}$)~$M_\odot$ for $m_X \simeq 1$ ($3$)~keV, below Milky-Way-like galaxy sizes. Consequently, the WDM power spectrum, $P_{\rm WDM}(k)$, can be written in terms of that for CDM $P_{\rm CDM}(k)$ through a transfer function $T_{\rm WDM}(k)$ which accounts for the suppression on small scales \citep{Bode:2000gq}:

\begin{equation}
\label{eq:pwdm}
P_{\rm WDM}(k) = T^2_{\rm WDM}(k) \, P_{\rm CDM}(k).
\end{equation}
This transfer function has to be obtained from a numerical Boltzmann code in order to accurately account for free-streaming. A common fit is given by ~\cite{Bode:2000gq,Viel:2005qj}
\begin{equation}
\label{eq:twdm}
  T_{\rm WDM}(k) = (1+ (\alpha k)^{2\nu})^{-5/\nu} ~, 
\end{equation}

with $\nu=1.12$ and the breaking scale
\begin{equation}
  \alpha= 0.049 \left(\frac{{\rm keV}}{m_X}\right)^{1.11}\left(\frac{\Omega_X}{0.25}\right)^{0.11}\left(\frac{h}{0.7}\right)^{1.22} \, {\rm Mpc}/h ~,
\end{equation}
being $\Omega_X$ the dark matter density fraction over the critical density and $h$ the reduced Hubble constant. The dependence of the dimensionless power spectrum $\Delta^2(k) = \frac{1}{2 \pi^2} k^3 P(k)$ on the WDM particle mass $m_X$ is shown in Fig.~\ref{fig:delta} for several values of $m_X$. Notice that, the lighter the particle is, the smaller is the value of $k$ at which the suppression of the power spectrum appears.

\section{Simulations and their calibration}
\label{sec:simscal}

For modeling the formation of cosmic structure and the process of reionization we use numerical technology from the Cosmic Reionization On Computers (CROC) project \citep{Gnedin:2014uta,Gnedin:2014xta}. These simulations include a wide range of physical effects which are relevant for modeling accurately the cosmic reionization, such as dynamics of dark matter and cosmic gas, star formation and stellar feedback, non-equilibrium ionization of hydrogen and helium, fully coupled 3D radiative
transfer, radiation-field-dependent cooling and heating functions for the metal enriched cosmic gas, etc \citep[see][for mode details]{Gnedin:2014uta}.

Since the primary effect of Warm Dark Matter is on small spatial and mass scales, in this paper we only use simulations in computational boxes of $20h^{-1}\dim{Mpc}$ on a side. As our results demonstrate, it would be simply a waste of computational resources to use larger boxes. Specifically, we use the simulations from the "Caiman" series introduced in \citet{Gnedin:2017gbf}. The primary advantage of these new series over the original simulations from \citet{Gnedin:2014uta} is that the new runs include corrections for weak numerical convergence \citep{Gnedin:2016g} and, hence, deliver numerically converged results. Specifically, we use Planck cosmology; mass resolution of our simulations is $M_1=7\times10^6\Msun$ and the spatial resolution is kept fixed in physical units to $100\dim{pc}$, just as in \citet{Gnedin:2017gbf}.

We use three sets of simulations in this paper. The pure CDM set (the same one used in \citet{Gnedin:2017gbf}) includes six independent realizations (labeled A-F) that are combined in computing averaged quantities to approximate true averages over the infinite universe. The second set include a single realization (A) of a WDM model with all physical and numerical parameters kept identical to the CDM case. Finally, the third set includes four out of six independent realizations (A,B,D,E) with the WDM model and star formation efficiency increased by a factor of 1.5 (or, equivalently, the gas depletion time decreased by a factor 1.5) compared to the CDM case. The particular choice of these four realizations is motivated by the property that the average over them in the pure CDM simulation set for all the quantities considered in this paper is virtually identical to analogous averages over all six realizations. Hence, these four realizations serve as a good approximation to the average universe.

\begin{figure}
\includegraphics[width=\hsize]{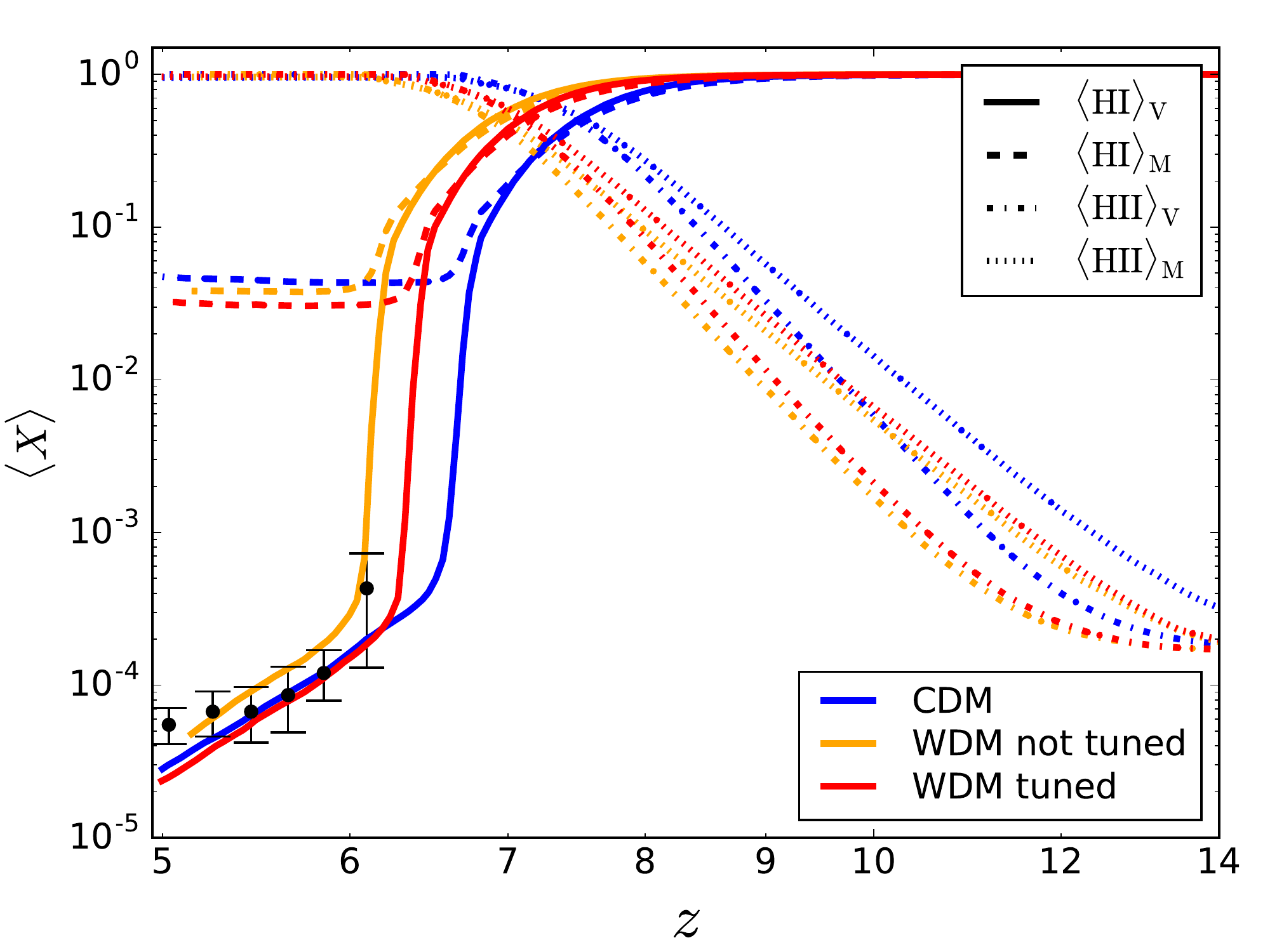}
\caption{Ionization histories for a single realization (A) of the CDM model (blue lines), the WDM model with identical physical and numerical parameters ("not \emph{tuned} WDM", orange lines) and the WDM model with 50\% higher star formation efficiency ("\emph{tuned} WDM", red lines). Solid (dotted-dashed) lines correspond to volume-weighted HI (HII) fraction, while dashed (dotted) lines correspond to mass-weighted HI (HII) fraction. The measurements are from \citet{Fan:2005es}. \label{fig:xhz}}
\end{figure}

The reason for adjusting the star formation efficiency is given in Figure \ref{fig:xhz}. It shows ionization histories for the first realization (A) for the three simulations sets we consider. The original CROC CDM simulations were calibrated against the diverse observational data to ensure the best achievable (not necessarily good) match. In particular, they match the observed mean opacity in the spectra of high redshift quasars from \citet{Fan:2005es}. The WDM model with identical physical and numerical parameters does not match these data, because the lack of small scale power reduces star formation and, hence, ionizing emissivity of dwarf galaxies. But parameters like star formation efficiency and ionizing emissivity are phenomenological adjustable parameters, they can not yet be constrained from the first principles. Hence, the WDM simulations need to be re-calibrated again to ensure the best possible match with the data in order to be a viable theoretical model. 

We have performed such re-calibration, and it turns out that it is only sufficient to increase the star formation efficiency by 50\%, while keeping fixed all other physical and numerical parameters to the previously calibrated values. Ionization history for the \emph{tuned} WDM model is also shown in Fig.\ \ref{fig:xhz} with the red lines. As one can see, it offers an identically good fit to \citet{Fan:2005es} data. We shall elaborate further on this point in our Results section. 

In the rest of the paper we will show that such a \emph{tuned} WDM model also fits some (but not all) observational constraints as well as the best-calibrated CDM model. Hence, hereafter we will only compare the CDM simulation set and the simulations set of the re-calibrated WDM model; we will drop the qualification \emph{tuned} for clarity, since the proper comparison between the two models is for their best calibrated realizations, not between models with the same physical and numerical parameters (in which case one of them would not be properly calibrated).

A particular challenge in modeling Warm or Hot Dark Matter with simulations is artificial numerical fragmentation, which has been extensively discussed in the literature \citep{Wang:2007he,Banerjee:2016zaa,Schneider:2013ria, Angulo:2013sza}. Shot noise due to initial random thermal velocities can seed unphysical overdensities at small scales, which lead to the formation of spurious structures as the simulation evolves. As a result, there is an unphysical over-abundance of low-mass halos, just the opposite from what is expected for a WDM model.

\begin{figure}
\includegraphics[width=\hsize]{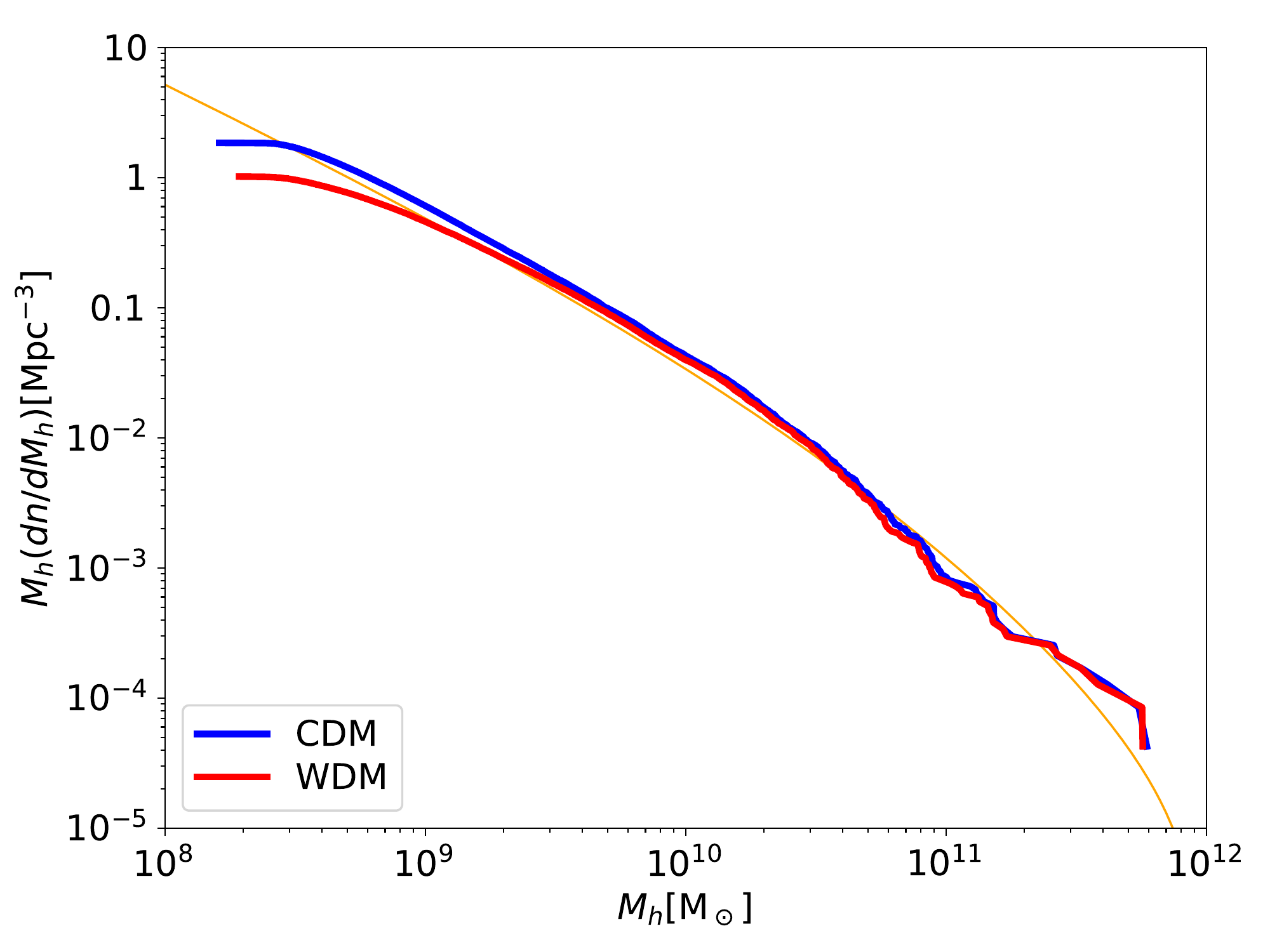}
\caption{Halo mass function for the CDM (blue) and WDM (red) case as well as theoretical expectation (orange) using \citet{Tinker:2008tkk} fitting formula. We avoid artificial numerical fragmentation, typical of WDM simulations, due to our particular refinement criterion.\label{fig:hmf}}
\end{figure}

There exist several numerical approaches for suppressing this artificial fragmentation \citep{Angulo:2013sza,Hobbs:2015dda}. In this work we adopt an alternative, simpler approach that is sufficient for our purpose, but may not be applicable to higher mass resolution simulations. CROC simulations use the Adaptive Refinement Tree (ART) code \citep{Kravtsov:1999,Kravtsov:2001ac,Rudd:2007zx} that implements an Adaptive Mesh Refinement technique. The mesh refinement and de-refinement strategies is essentially user-definable; in CROC simulations refinement is implemented in a quasi-Lagrangian manner, striving to keep the mass per cell approximately constant, but the refinement is performed only on the gas density and not on the dark matter density. Since artificial numerical fragmentation typically produces halos with masses too small to contain any appreciable amount of baryons \cite{Lovell:2013ola,Okamoto:2008sn}, the computational mesh (and, hence all gravity calculations) is not refined at the potential locations of such halos, and thus they simply do not form in CROC simulations. This is apparent from Figure \ref{fig:hmf}, that shows halo mass functions for the CDM and WDM simulations with identical numerical parameters. 

It is important to note here that the simulations presented in Fig.\ \ref{fig:hmf} have limited mass resolution, resolving halos down to $M_h\approx3\times10^8\Msun$ (the mass at which the simulated CDM halo mass function deviates from the theoretical expectation). It is possible that for higher resolution simulations artificial numerical fragmentation would appear even with our choice of refinement criteria.

\section{Results}
\label{sec:res}
\subsection{Galaxy luminosity functions}

\begin{figure}
\includegraphics[width=\hsize]{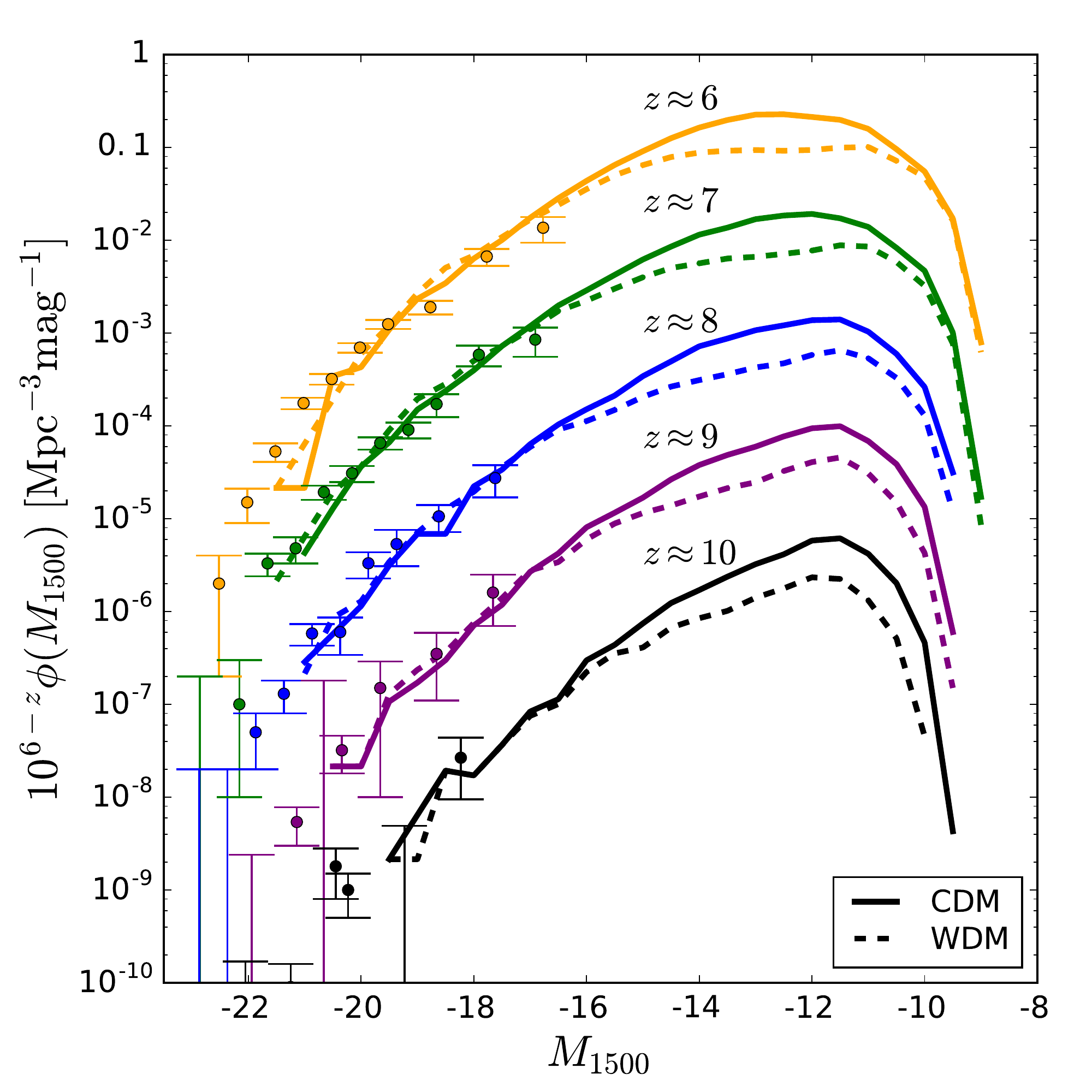}
\caption{Galaxy UV luminosity functions at five different redshifts for CDM (solid) and WDM (dashed) simulations. Points with error bars from \cite{Bouwens:2014fua}.\label{fig:lfz}}
\end{figure}

Galaxy UV luminosity functions represent one of the best observables to constrain the properties of reionization sources. Current HST observations have managed to measure UV magnitudes $\sim -15$ at $z\la8$ and reach redshifts $z\sim 10$ \citep{Bouwens:2014fua}. Despite their high sensitivity, the current observations are still insensitive to the faint end of the  UV luminosity function for the range of galaxy luminosities affected by the nature of dark matter. In \cite{Gnedin:2016vdz} it was shown that simulations  agree well with current measurements, and that the usual Schechter fit can provide a good description of the luminosity functions over the currently measured range of luminosities.

Figure \ref{fig:lfz} depicts one of the main results of this study. We plot the simulated UV galaxy luminosity functions for $z=$6, 7, 8, 9, and 10, together with current observations from \citet{Bouwens:2014fua}. The solid lines depict the CDM scenario, while the dashed lines show the predictions for a WDM particle with a mass $m_X=3$~keV. Notice that both models match well each other \emph{and} the existing measurements; however, at lower luminosities the WDM model predicts fewer galaxies, as expected. It is worth highlighting that \textit{(a)} there is a difference between the CDM and WDM cosmologies for magnitudes $\sim -14$ and further beyond; and \textit{(b)} this difference is almost redshift independent. 

The current observations have not yet reached this crucial region, and existing data do not allow us to distinguish among the CDM and WDM models. However, the future JWST observations will be able to probe galaxy luminosity functions at redshifts $z\sim 8-10$ down to magnitudes $\sim -13$ by taking advantage of gravitational lensing effects. Whether these observations will be able to actually constraint the nature of dark matter is, however, unclear, due to inherent theoretical uncertainty in predicting galaxy luminosity functions at these faint magnitudes. We expand on this subject further in the Conclusions.

Another feature of modeled galaxy luminosity function in Fig.\ \ref{fig:lfz} is the steep turnover at $M_{1500}\approx -10$. This turnover is only weakly sensitive to the nature of dark matter and is highly model dependent. It is caused by the primary assumption made in CROC simulations that stars form predominantly in molecular gas. Since halos with masses below about $10^8\Msun$ contain little molecular gas, they contribute respectively little to star formation (and, hence to UV luminosity). This result, we emphasize again, is a theoretical assumption, not a known fact. Several recent large computational projects achieved similar level of precision in matching observational constraints to the CROC projects, and they all predict the turnover to occur at widely varying luminosities \citep{Bouwens:2017boi}. Hence, the location and the shape of the turnover cannot be considered as a reliable theoretical prediction and will have to be constrained by future observations.

\subsection{Ionization history}

The ionization history of the universe may also offer a tool to constrain the dark matter nature. The ionization history is given by the volume filling factor of the ionized hydrogen $\qi$, whose evolution follows the reionization equation of \citet{Madau:1998cd}:
\begin{equation}
  \frac{d\qi}{dt} = \frac{\dot{n}_{\rm ion}}{n_{\rm H}} - \frac{\qi}{\bar{t}_{\rm rec}},
  \label{eq:mhr}
\end{equation}
where $\dot{n}_{\rm ion}$ is the globally averaged rate of production of hydrogen ionizing photons, which can be related with the star formation rate $\dot{\rho}_*$ through \citep{Robertson:2015uda}
\begin{equation}
  \dot{n}_{\rm ion} = f_{\rm esc} \xi_{\rm ion} \dot{\rho}_*,
  \label{eq:nion}
\end{equation}
where $ f_{\rm esc}$ is the effective escape fraction of ionizing photons and $\xi_{\rm ion}$ the ionizing photon production efficiency per unit star formation rate.

As previously stated, in WDM cosmologies, free-streaming of dark matter particles suppresses small-scale fluctuations, delaying structure formation and, therefore, the onset of the overall reionization process. The effect will be more pronounced for smaller WDM masses (see Fig.~\ref{fig:delta}). However, after a quick inspection of Eq.~(\ref{eq:nion}), one can notice that the eventual smaller ionized fraction in the presence of a WDM particle could be easily accommodated by an increase of the star formation rate/effective escape fraction of ionizing photons, leaving unchanged the volume filling factor. Therefore, as anticipated in Sec.~\ref{sec:simscal} and illustrated in Fig.~\ref{fig:xhz}, it is always possible to match the ionization history of a given WDM cosmology to that expected in the standard CDM picture and be perfectly consistent with SDSS high redshift quasars observations at $5.74 < z < 6.42$ \citep{Fan:2005es}. 

While Fig.\ \ref{fig:xhz} shows that some differences between our CDM and WDM models remain even after tuning, it is unlikely to be a detectable difference, as we have only tuned the star formation efficiency. There exist many more parameters that can also be tuned in the simulations (cosmological parameters, feedback models, ionizing efficiency, etc) to achieve an even better agreement between CDM and WDM models.

\subsection{Post-reionization IGM}

\begin{figure}
\includegraphics[width=\hsize]{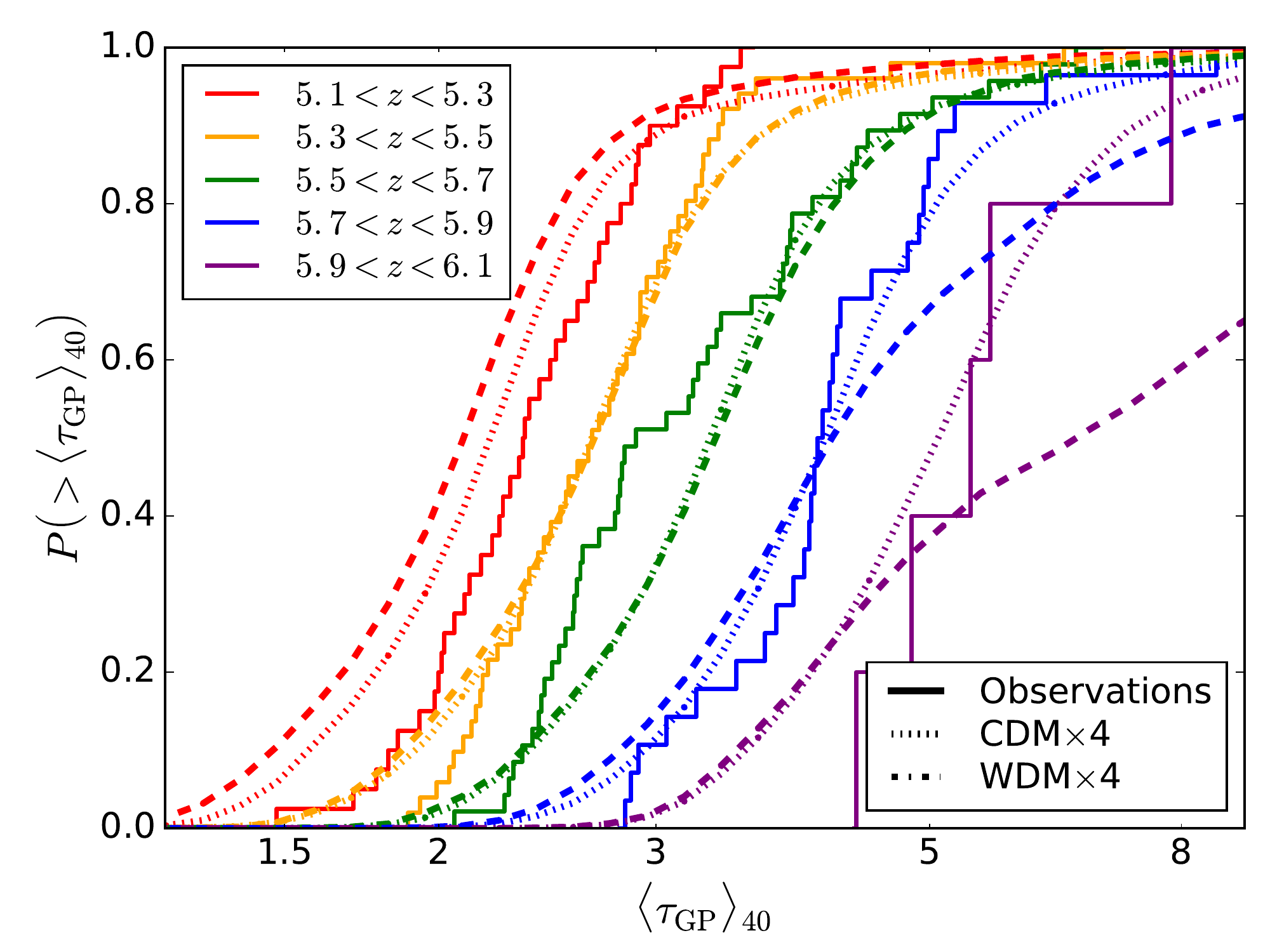}
\caption{Probability distribution functions for the effective IGM opacity in $40h^{-1}$Mpc sightline segments from our simulations. Solid lines correspond to CDM, dashed to WDM. Observational data are from \citet{Becker:2014oga} and \citet{Gnedin:2016wrw}. Later reionization in the WDM model is easily detectable by the data.\label{fig:ptau}}
\end{figure}

Galaxy UV luminosity functions offer a strong constraint on the sources of ionizing photons during reionization. Another strong observational constraint on the final stages of reionization and its immediate aftermath comes from the observations of absorption spectra of high redshift quasars. With the sample of quasars discovered at $z>6$ exceeding 100 \citep{Banados:2016bvd}, one can not only accurately measure the mean Gunn-Petrson opacity (as shown in Fig.\ \ref{fig:xhz}), but also determine their full PDFs \citep{Becker:2014oga}. 

Resonant scattering in the Ly-$\alpha$ line of (remaining after reionization residual) neutral hydrogen depletes the observed flux from a distant quasar by a factor $e^{-\tau}$, being $\tau$ the Gunn-Peterson optical depth. The value of $\tau$ is highly (quadratically) sensitive to the distribution of neutral hydrogen $n_{\rm{HI}}$ along a given line of sight. In order to have a statistical description of the Ly-$\alpha$ opacity, one usually averages the normalized flux over a number of lines of sight $\langle F \rangle$. It is customary to define an effective averaged optical depth $\tau_{eff}$ as
\begin{equation}
\tau_{eff} = -\textrm{ln} \langle F \rangle_L,
\end{equation}
where $L$ is the distance along the line of sight over which the averaging is performed.

We compute from our simulations the probability distribution function (PDF) of $\tau_{eff}$ and its cumulative distribution $P( \leq \tau_{eff})$, i.e., the probability of having an optical depth lower than $\tau_{eff}$. We test our predictions against the measurements obtained by \citet{Becker:2014oga} in Figure \ref{fig:ptau}, with averaging performed over sightline segments of comoving length $L=40h^{-1}\dim{Mpc}$, as in \citet{Gnedin:2017gbf}, who demonstrated that this segment length is reliably modeled by CROC simulations. Since the optical depth increases with redshift, the cumulative probability distribution is shifted to higher values of $\tau$, as $z$ increases. 

The effect of the WDM free-streaming nature on the Gunn-Peterson optical depth is easy to understand: since reionization is delayed in the WDM case, the higher density (and, hence, higher optical depth) regions get ionized later \citep{Miralda:2000mhr,Kaurov:2016k}, extending the high $\tau_{eff}$ tail towards larger values. The lower density regions , which have been ionized early enough, follow similar ionization states in both CDM and WDM models. Even with present data, the difference between the two models is easily detectable.

Both models also fail to match observations at $z<5.3$. This discrepancy was noted in the first CROC papers \citep{Gnedin:2014xta,Gnedin:2014uta} and was interpreted as a failure of the adopted stellar feedback model.

\subsection{Redshifted 21 cm emission}

Another potential strong constraint on the process of reionization is offered by the redshifted 21 cm emission from neutral hydrogen. While the idea to use that emission as a probe of reionization has been around for a long time (see \citet{21cm:pl12} for a recent review), only recently interesting observational constraints have been placed by Murchison Widefield Array \citep[MWA; ][]{21cm:dlw14,21cm:dnw15} and Donald C.\ Backer Precision Array for Probing the Epoch of Reionization \citep[PAPER][]{21cm:pla14,21cm:jpp15,21cm:pap15,21cm:apz15}).

\begin{figure}
\includegraphics[width=\hsize]{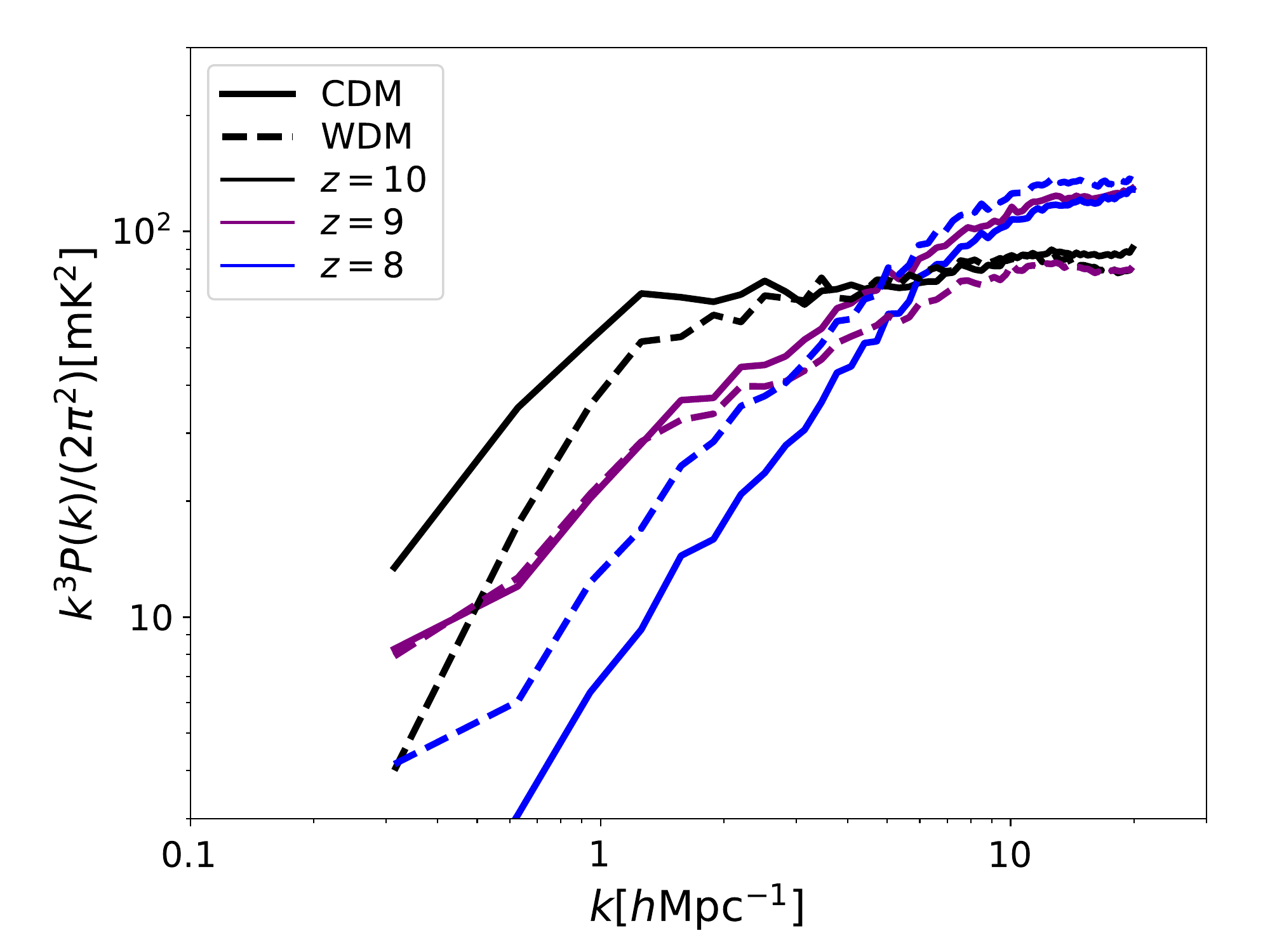}
\caption{Power spectra of redshifted 21 cm emission for CDM (solid) and WDM (dashed) models, at several redshifts where the signal is sufficiently high. \label{fig:ps21}}
\end{figure}

In Figure \ref{fig:ps21} we show simulated power spectra of redshifted 21 cm emission (the currently most favorable statistical probe of the reionization signal) at several redshifts. The differences between the CDM and WDM models are non-trivial, but are probably not large enough to be detectable by the first generation of experiments.

\section{Conclusions}

Warm Dark Matter scenarios with free-streaming scales of $\sim$Mpc, while giving excellent predictions for the measurements of the large scale structure of the universe, also offer an appealing solution to several unresolved difficulties in the standard CDM paradigm, such as the so-called "small-scale crisis" (missing satellite, too-big-to-fail, and core-cusp problems) and the controversial detection of a monochromatic line at $3.56$~keV in X-ray data. Furthermore, particle physics renders many well-motivated theoretical scenarios in which WDM candidates arise naturally (e.g. sterile neutrinos). 

It is therefore timely to explore what the reionization history of the universe can currently tell about the nature of dark matter, specifically focusing on a WDM particle mass of $\sim 3$~keV, the present lower bound from CMB and Ly-$\alpha$ forest power spectra. The formation of cosmic structure and the process of reionization are sufficiently realistically simulated via the Cosmic Reionization On Computers (CROC) project \citep{Gnedin:2014uta,Gnedin:2014xta}. These simulations account for the dynamics of dark matter and cosmic gas, star formation and stellar feedback, non-equilibrium ionization of hydrogen and helium, fully coupled 3D radiative transfer, and radiation-field-dependent cooling and heating functions for the metal enriched cosmic gas. Three sets of simulations have been produced: the canonical CDM cosmology case, a WDM model with identical parameters to the CDM case, and a WDM model in which the star formation rate is increased by a factor of 1.5. 

We have shown that the lack of small scale power in WDM cosmology (relative to the equivalent CDM case) due to the non-negligible free-streaming length of WDM particle considerably delays the reionization processes. However, a higher star formation efficiency (or, equivalently, a lower gas depletion time) compensates for the WDM-suppressed small-scale structure, leading to nearly identical (within the currently observationally constrained range) galaxy luminosity functions in the CDM and WDM cases. Such tuning of the star formation efficiency in our simulations is perfectly justified, as this quantity still remains a parameter, adjustable within reasonable bounds, that currently can be derived neither from the first principles nor from observational constraints. 

A so tuned WDM model never-the-less underpredicts (as compared to the pure CDM case) the abundance of galaxies with UV magnitudes below about -15, the range of galactic luminosities well within reach of the soon-to-be-launched JWST telescope (which is expected to measure the faint end of the galaxy UV luminosity function down to M$_{\rm{UV}} \simeq -13$). The WDM model also differs significantly from the CDM case in the predicted redshifted 21cm emission. 

Clustering and luminosity functions of Ly-$\alpha$ emitters~\citep{2017arXiv170501222K,Ouchi:2017cqe} may offer another possible test of WDM scenarios. In practice, such a test strongly relies on comparisons with analytical models and therefore is a highly model-dependent probe of reionization. 

However, by far the most significant signature of delayed reionization in the WDM case is the distribution of Gunn-Peterson optical depth in the spectra of high-redshift quasars - perhaps, the most constraining today observational data on cosmic reionization. Both the Planck value of the Thomson optical depth~\citep{Adam:2016hgk,Aghanim:2016yuo} and the ionization state of low density IGM significantly constrain reionization history with the existing data. Both those constraints push reionization to lower redshifts. This push is, however, resisted by the observational constraints on the ionization state of higher density gas (i.e. higher $\tau_{eff}$ tail of Gunn-Peterson optical depth distributions). Hence, the overall shape of PDFs of $\tau_{eff}$ is strongly constraining. In particular, our WDM model fails that observational test by a large margin even after adjusting the star formation efficiency.

Despite all these observational signatures, the question of whether the nature of dark matter can be constrained by observational tests of cosmic reionization remains open. In our simulations we tuned the star formation efficiency to reproduce the observed UV luminosity functions in both CDM and WDM models, but galaxy formation simulations include much more uncertainty that just one parameter - the full model of star formation and stellar feedback is largely unconstrained. For example, it is possible that some other change to the stellar feedback model adopted in the simulations may reduce the feedback strength and make the WDM model more consistent with observations. However, such a change must be highly non-trivial, since the stellar feedback in CROC simulations is already too weak - it fails to reduce star formation by $z=5$ sufficiently to match the Gunn-Peterson data for $z<5.3$.

Thus, whether reionization observables can actually be used to constraint the nature of dark matter ultimately depends on the ability of future simulations to predict the shape of galaxy luminosity function precisely - a strict requirement, only possible if the physics of star formation and stellar feedback in these dwarf, high redshift galaxies can be constrained sufficiently accurately. The field of galaxy formation is seeing lighting speed progress in modeling stellar feedback, but the required precision in modeling galaxy luminosity functions remains a high order, and only future will show whether that order can be fulfilled.

\section*{acknowledgements}
OM and PVD would like to thank the Fermilab Theoretical Physics Department for hospitality. OM and PVD are supported by PROMETEO II/2014/050, by the Spanish Grants SEV-2014-0398 and FPA2014--57816-P of MINECO and by the European Union's Horizon 2020 research and innovation program under the Marie Sk\l odowska-Curie grant agreements No. 690575 and 674896.   

This manuscript has been authored in part by Fermi Research Alliance, LLC under Contract No. DE-AC02-07CH11359 with the U.S. Department of Energy, Office of Science, Office of High Energy Physics. The United States Government retains and the publisher, by accepting the article for publication, acknowledges that the United States Government retains a non-exclusive, paid-up, irrevocable, world-wide license to publish or reproduce the published form of this manuscript, or allow others to do so, for United States Government purposes. This research used resources of the Argonne Leadership Computing Facility, which is a DOE Office of Science User Facility supported under Contract DE-AC02-06CH11357. An award of computer time was provided by the Innovative and Novel Computational Impact on Theory and Experiment (INCITE) program. This research is also part of the Blue Waters sustained-petascale computing project, which is supported by the National Science Foundation (awards OCI-0725070 and ACI-1238993) and the state of Illinois. Blue Waters is a joint effort of the University of Illinois at Urbana-Champaign and its National Center for Supercomputing Applications. This work made extensive use of the NASA Astrophysics Data System and {\tt arXiv.org} preprint server.

\bibliographystyle{apj}
\bibliography{biblio}

\begin{thebibliography}{85}
\expandafter\ifx\csname natexlab\endcsname\relax\def\natexlab#1{#1}\fi

\bibitem[{Adam {et~al.}(2016)}]{Adam:2016hgk}
Adam, R. {et~al.} 2016, Astron. Astrophys., 596, A108

\bibitem[{Ade {et~al.}(2016)}]{Ade:2015xua}
Ade, P. A.~R. {et~al.} 2016, Astron. Astrophys., 594, A13

\bibitem[{Aghanim {et~al.}(2016)}]{Aghanim:2016yuo}
Aghanim, N. {et~al.} 2016, Astron. Astrophys., 596, A107

\bibitem[{Alam {et~al.}(2016)}]{Alam:2016hwk}
Alam, S. {et~al.} 2016, Submitted to: Mon. Not. Roy. Astron. Soc.

\bibitem[{{Ali} {et~al.}(2015){Ali}, {Parsons}, {Zheng}, {Pober}, {Liu},
  {Aguirre}, {Bradley}, {Bernardi}, {Carilli}, {Cheng}, {DeBoer}, {Dexter},
  {Grobbelaar}, {Horrell}, {Jacobs}, {Klima}, {MacMahon}, {Maree}, {Moore},
  {Razavi}, {Stefan}, {Walbrugh}, \& {Walker}}]{21cm:apz15}
{Ali}, Z.~S., {Parsons}, A.~R., {Zheng}, H., {Pober}, J.~C., {Liu}, A.,
  {Aguirre}, J.~E., {Bradley}, R.~F., {Bernardi}, G., {Carilli}, C.~L.,
  {Cheng}, C., {DeBoer}, D.~R., {Dexter}, M.~R., {Grobbelaar}, J., {Horrell},
  J., {Jacobs}, D.~C., {Klima}, P., {MacMahon}, D.~H.~E., {Maree}, M., {Moore},
  D.~F., {Razavi}, N., {Stefan}, I.~I., {Walbrugh}, W.~P., \& {Walker}, A.
  2015, \apj, 809, 61

\bibitem[{Angulo {et~al.}(2013)Angulo, Hahn, \& Abel}]{Angulo:2013sza}
Angulo, R.~E., Hahn, O., \& Abel, T. 2013, Mon. Not. Roy. Astron. Soc., 434,
  3337

\bibitem[{{Ba{\~n}ados} {et~al.}(2016){Ba{\~n}ados}, {Venemans}, {Decarli},
  {Farina}, {Mazzucchelli}, {Walter}, {Fan}, {Stern}, {Schlafly}, {Chambers},
  {Rix}, {Jiang}, {McGreer}, {Simcoe}, {Wang}, {Yang}, {Morganson}, {De Rosa},
  {Greiner}, {Balokovi{\'c}}, {Burgett}, {Cooper}, {Draper}, {Flewelling},
  {Hodapp}, {Jun}, {Kaiser}, {Kudritzki}, {Magnier}, {Metcalfe}, {Miller},
  {Schindler}, {Tonry}, {Wainscoat}, {Waters}, \& {Yang}}]{Banados:2016bvd}
{Ba{\~n}ados}, E., {Venemans}, B.~P., {Decarli}, R., {Farina}, E.~P.,
  {Mazzucchelli}, C., {Walter}, F., {Fan}, X., {Stern}, D., {Schlafly}, E.,
  {Chambers}, K.~C., {Rix}, H.-W., {Jiang}, L., {McGreer}, I., {Simcoe}, R.,
  {Wang}, F., {Yang}, J., {Morganson}, E., {De Rosa}, G., {Greiner}, J.,
  {Balokovi{\'c}}, M., {Burgett}, W.~S., {Cooper}, T., {Draper}, P.~W.,
  {Flewelling}, H., {Hodapp}, K.~W., {Jun}, H.~D., {Kaiser}, N., {Kudritzki},
  R.-P., {Magnier}, E.~A., {Metcalfe}, N., {Miller}, D., {Schindler}, J.-T.,
  {Tonry}, J.~L., {Wainscoat}, R.~J., {Waters}, C., \& {Yang}, Q. 2016, \apjs,
  227, 11

\bibitem[{Banerjee \& Dalal(2016)}]{Banerjee:2016zaa}
Banerjee, A. \& Dalal, N. 2016, JCAP, 1611, 015

\bibitem[{Barkana {et~al.}(2001)Barkana, Haiman, \& Ostriker}]{Barkana:2001gr}
Barkana, R., Haiman, Z., \& Ostriker, J.~P. 2001, Astrophys. J., 558, 482

\bibitem[{Baur {et~al.}(2016)Baur, Palanque-Delabrouille, Y{\`e}che,
  Magneville, \& Viel}]{Baur:2015jsy}
Baur, J., Palanque-Delabrouille, N., Y{\`e}che, C., Magneville, C., \& Viel, M.
  2016, JCAP, 1608, 012

\bibitem[{Becker {et~al.}(2015)Becker, Bolton, Madau, Pettini, Ryan-Weber, \&
  Venemans}]{Becker:2014oga}
Becker, G.~D., Bolton, J.~S., Madau, P., Pettini, M., Ryan-Weber, E.~V., \&
  Venemans, B.~P. 2015, Mon. Not. Roy. Astron. Soc., 447, 3402

\bibitem[{Bergstrom(2012)}]{Bergstrom:2012fi}
Bergstrom, L. 2012, Annalen Phys., 524, 479

\bibitem[{Bertone {et~al.}(2005)Bertone, Hooper, \& Silk}]{Bertone:2004pz}
Bertone, G., Hooper, D., \& Silk, J. 2005, Phys. Rept., 405, 279

\bibitem[{Bode {et~al.}(2001)Bode, Ostriker, \& Turok}]{Bode:2000gq}
Bode, P., Ostriker, J.~P., \& Turok, N. 2001, Astrophys. J., 556, 93

\bibitem[{Bose {et~al.}(2016)Bose, Frenk, Hou, Lacey, \& Lovell}]{Bose:2016hlz}
Bose, S., Frenk, C.~S., Hou, J., Lacey, C.~G., \& Lovell, M.~R. 2016, Mon. Not.
  Roy. Astron. Soc., 463, 3848

\bibitem[{Bose {et~al.}(2017)Bose, Hellwing, Frenk, Jenkins, Lovell, Helly, Li,
  Gonzalez-Perez, \& Gao}]{Bose:2016irl}
Bose, S., Hellwing, W.~A., Frenk, C.~S., Jenkins, A., Lovell, M.~R., Helly,
  J.~C., Li, B., Gonzalez-Perez, V., \& Gao, L. 2017, Mon. Not. Roy. Astron.
  Soc., 464, 4520

\bibitem[{{Bouwens} {et~al.}(2017){Bouwens}, {Oesch}, {Illingworth}, {Ellis},
  \& {Stefanon}}]{Bouwens:2017boi}
{Bouwens}, R.~J., {Oesch}, P.~A., {Illingworth}, G.~D., {Ellis}, R.~S., \&
  {Stefanon}, M. 2017, \apj, 843, 129

\bibitem[{Bouwens {et~al.}(2015)}]{Bouwens:2014fua}
Bouwens, R.~J. {et~al.} 2015, Astrophys. J., 803, 34

\bibitem[{Boyarsky {et~al.}(2009)Boyarsky, Lesgourgues, Ruchayskiy, \&
  Viel}]{Boyarsky:2008xj}
Boyarsky, A., Lesgourgues, J., Ruchayskiy, O., \& Viel, M. 2009, JCAP, 0905,
  012

\bibitem[{Boyarsky {et~al.}(2014)Boyarsky, Ruchayskiy, Iakubovskyi, \&
  Franse}]{Boyarsky:2014jta}
Boyarsky, A., Ruchayskiy, O., Iakubovskyi, D., \& Franse, J. 2014, Phys. Rev.
  Lett., 113, 251301

\bibitem[{Boylan-Kolchin {et~al.}(2012)Boylan-Kolchin, Bullock, \&
  Kaplinghat}]{BoylanKolchin:2011dk}
Boylan-Kolchin, M., Bullock, J.~S., \& Kaplinghat, M. 2012, Mon. Not. Roy.
  Astron. Soc., 422, 1203

\bibitem[{Bulbul {et~al.}(2014)}]{Bulbul:2014sua}
Bulbul, E. {et~al.} 2014, Astrophys. J., 789, 13

\bibitem[{Bullock \& Boylan-Kolchin(2017)}]{Bullock:2017xww}
Bullock, J.~S. \& Boylan-Kolchin, M. 2017, Ann. Rev. Astron. Astrophys., 55,
  343

\bibitem[{Corasaniti {et~al.}(2017)Corasaniti, Agarwal, Marsh, \&
  Das}]{Corasaniti:2016epp}
Corasaniti, P.~S., Agarwal, S., Marsh, D. J.~E., \& Das, S. 2017, Phys. Rev.,
  D95, 083512

\bibitem[{Dayal {et~al.}(2017)Dayal, Choudhury, Bromm, \&
  Pacucci}]{Dayal:2015vca}
Dayal, P., Choudhury, T.~R., Bromm, V., \& Pacucci, F. 2017, Astrophys. J.,
  836, 16

\bibitem[{{Dillon} {et~al.}(2014){Dillon}, {Liu}, {Williams}, {Hewitt},
  {Tegmark}, {Morgan}, {Levine}, {Morales}, {Tingay}, {Bernardi}, {Bowman},
  {Briggs}, {Cappallo}, {Emrich}, {Mitchell}, {Oberoi}, {Prabu}, {Wayth}, \&
  {Webster}}]{21cm:dlw14}
{Dillon}, J.~S., {Liu}, A., {Williams}, C.~L., {Hewitt}, J.~N., {Tegmark}, M.,
  {Morgan}, E.~H., {Levine}, A.~M., {Morales}, M.~F., {Tingay}, S.~J.,
  {Bernardi}, G., {Bowman}, J.~D., {Briggs}, F.~H., {Cappallo}, R.~C.,
  {Emrich}, D., {Mitchell}, D.~A., {Oberoi}, D., {Prabu}, T., {Wayth}, R., \&
  {Webster}, R.~L. 2014, \prd, 89, 023002

\bibitem[{{Dillon} {et~al.}(2015){Dillon}, {Neben}, {Hewitt}, {Tegmark},
  {Barry}, {Beardsley}, {Bowman}, {Briggs}, {Carroll}, {de Oliveira-Costa},
  {Ewall-Wice}, {Feng}, {Greenhill}, {Hazelton}, {Hernquist}, {Hurley-Walker},
  {Jacobs}, {Kim}, {Kittiwisit}, {Lenc}, {Line}, {Loeb}, {McKinley},
  {Mitchell}, {Morales}, {Offringa}, {Paul}, {Pindor}, {Pober}, {Procopio},
  {Riding}, {Sethi}, {Shankar}, {Subrahmanyan}, {Sullivan}, {Thyagarajan},
  {Tingay}, {Trott}, {Wayth}, {Webster}, {Wyithe}, {Bernardi}, {Cappallo},
  {Deshpande}, {Johnston-Hollitt}, {Kaplan}, {Lonsdale}, {McWhirter}, {Morgan},
  {Oberoi}, {Ord}, {Prabu}, {Srivani}, {Williams}, \& {Williams}}]{21cm:dnw15}
{Dillon}, J.~S., {Neben}, A.~R., {Hewitt}, J.~N., {Tegmark}, M., {Barry}, N.,
  {Beardsley}, A.~P., {Bowman}, J.~D., {Briggs}, F., {Carroll}, P., {de
  Oliveira-Costa}, A., {Ewall-Wice}, A., {Feng}, L., {Greenhill}, L.~J.,
  {Hazelton}, B.~J., {Hernquist}, L., {Hurley-Walker}, N., {Jacobs}, D.~C.,
  {Kim}, H.~S., {Kittiwisit}, P., {Lenc}, E., {Line}, J., {Loeb}, A.,
  {McKinley}, B., {Mitchell}, D.~A., {Morales}, M.~F., {Offringa}, A.~R.,
  {Paul}, S., {Pindor}, B., {Pober}, J.~C., {Procopio}, P., {Riding}, J.,
  {Sethi}, S., {Shankar}, N.~U., {Subrahmanyan}, R., {Sullivan}, I.,
  {Thyagarajan}, N., {Tingay}, S.~J., {Trott}, C., {Wayth}, R.~B., {Webster},
  R.~L., {Wyithe}, S., {Bernardi}, G., {Cappallo}, R.~J., {Deshpande}, A.~A.,
  {Johnston-Hollitt}, M., {Kaplan}, D.~L., {Lonsdale}, C.~J., {McWhirter},
  S.~R., {Morgan}, E., {Oberoi}, D., {Ord}, S.~M., {Prabu}, T., {Srivani},
  K.~S., {Williams}, A., \& {Williams}, C.~L. 2015, \prd, 91, 123011

\bibitem[{Fan {et~al.}(2006)}]{Fan:2005es}
Fan, X.-H. {et~al.} 2006, Astron. J., 132, 117

\bibitem[{Fattahi {et~al.}(2016)}]{Fattahi:2016nld}
Fattahi, A. {et~al.} 2016

\bibitem[{Gardner {et~al.}(2006)}]{Gardner:2006ky}
Gardner, J.~P. {et~al.} 2006, Space Sci. Rev., 123, 485

\bibitem[{Gnedin(2014)}]{Gnedin:2014uta}
Gnedin, N.~Y. 2014, Astrophys. J., 793, 29

\bibitem[{{Gnedin}(2016)}]{Gnedin:2016g}
{Gnedin}, N.~Y. 2016, \apj, 821, 50

\bibitem[{Gnedin(2016)}]{Gnedin:2016vdz}
Gnedin, N.~Y. 2016, Astrophys. J., 825, L17

\bibitem[{{Gnedin} {et~al.}(2017){Gnedin}, {Becker}, \& {Fan}}]{Gnedin:2017gbf}
{Gnedin}, N.~Y., {Becker}, G.~D., \& {Fan}, X. 2017, \apj, 841, 26

\bibitem[{Gnedin {et~al.}(2017)Gnedin, Becker, \& Fan}]{Gnedin:2016wrw}
Gnedin, N.~Y., Becker, G.~D., \& Fan, X. 2017, Astrophys. J., 841, 26

\bibitem[{Gnedin \& Kaurov(2014)}]{Gnedin:2014xta}
Gnedin, N.~Y. \& Kaurov, A.~A. 2014, Astrophys. J., 793, 30

\bibitem[{Hobbs {et~al.}(2016)Hobbs, Read, Agertz, Iannuzzi, \&
  Power}]{Hobbs:2015dda}
Hobbs, A., Read, J., Agertz, O., Iannuzzi, F., \& Power, C. 2016, Mon. Not.
  Roy. Astron. Soc., 458, 468

\bibitem[{Ir{\v{s}}i{\v{c}} {et~al.}(2017)}]{Irsic:2017ixq}
Ir{\v{s}}i{\v{c}}, V. {et~al.} 2017

\bibitem[{{Jacobs} {et~al.}(2015){Jacobs}, {Pober}, {Parsons}, {Aguirre},
  {Ali}, {Bowman}, {Bradley}, {Carilli}, {DeBoer}, {Dexter}, {Gugliucci},
  {Klima}, {Liu}, {MacMahon}, {Manley}, {Moore}, {Stefan}, \&
  {Walbrugh}}]{21cm:jpp15}
{Jacobs}, D.~C., {Pober}, J.~C., {Parsons}, A.~R., {Aguirre}, J.~E., {Ali},
  Z.~S., {Bowman}, J., {Bradley}, R.~F., {Carilli}, C.~L., {DeBoer}, D.~R.,
  {Dexter}, M.~R., {Gugliucci}, N.~E., {Klima}, P., {Liu}, A., {MacMahon},
  D.~H.~E., {Manley}, J.~R., {Moore}, D.~F., {Stefan}, I.~I., \& {Walbrugh},
  W.~P. 2015, \apj, 801, 51

\bibitem[{{Kaurov}(2016)}]{Kaurov:2016k}
{Kaurov}, A.~A. 2016, \apj, 831, 198

\bibitem[{Klypin {et~al.}(1999)Klypin, Kravtsov, Valenzuela, \&
  Prada}]{Klypin:1999uc}
Klypin, A.~A., Kravtsov, A.~V., Valenzuela, O., \& Prada, F. 1999, Astrophys.
  J., 522, 82

\bibitem[{Kolb \& Turner(1990)}]{Kolb:1990vq}
Kolb, E.~W. \& Turner, M.~S. 1990, Front. Phys., 69, 1

\bibitem[{{Konno} {et~al.}(2017){Konno}, {Ouchi}, {Shibuya}, {Ono},
  {Shimasaku}, {Taniguchi}, {Nagao}, {Kobayashi}, {Kajisawa}, {Kashikawa},
  {Inoue}, {Oguri}, {Furusawa}, {Goto}, {Harikane}, {Higuchi}, {Komiyama},
  {Kusakabe}, {Miyazaki}, {Nakajima}, \& {Wang}}]{2017arXiv170501222K}
{Konno}, A., {Ouchi}, M., {Shibuya}, T., {Ono}, Y., {Shimasaku}, K.,
  {Taniguchi}, Y., {Nagao}, T., {Kobayashi}, M.~A.~R., {Kajisawa}, M.,
  {Kashikawa}, N., {Inoue}, A.~K., {Oguri}, M., {Furusawa}, H., {Goto}, T.,
  {Harikane}, Y., {Higuchi}, R., {Komiyama}, Y., {Kusakabe}, H., {Miyazaki},
  S., {Nakajima}, K., \& {Wang}, S.-Y. 2017, ArXiv e-prints

\bibitem[{Kravtsov(1999)}]{Kravtsov:1999}
Kravtsov, A.~V. 1999, PhD thesis, New Mexico State University

\bibitem[{Kravtsov {et~al.}(2002)Kravtsov, Klypin, \&
  Hoffman}]{Kravtsov:2001ac}
Kravtsov, A.~V., Klypin, A.~A., \& Hoffman, Y. 2002, Astrophys. J., 571, 563

\bibitem[{Lapi \& Danese(2015)}]{Lapi:2015zea}
Lapi, A. \& Danese, L. 2015, JCAP, 1509, 003

\bibitem[{Lopez-Honorez {et~al.}(2017)Lopez-Honorez, Mena, Palomares-Ruiz, \&
  Domingo}]{Lopez-Honorez:2017csg}
Lopez-Honorez, L., Mena, O., Palomares-Ruiz, S., \& Domingo, P.~V. 2017

\bibitem[{Lovell {et~al.}(2014)Lovell, Frenk, Eke, Jenkins, Gao, \&
  Theuns}]{Lovell:2013ola}
Lovell, M.~R., Frenk, C.~S., Eke, V.~R., Jenkins, A., Gao, L., \& Theuns, T.
  2014, Mon. Not. Roy. Astron. Soc., 439, 300

\bibitem[{Lovell {et~al.}(2012)}]{Lovell:2011rd}
Lovell, M.~R. {et~al.} 2012, Mon. Not. Roy. Astron. Soc., 420, 2318

\bibitem[{Lovell {et~al.}(2016)}]{Lovell:2015psz}
---. 2016, Mon. Not. Roy. Astron. Soc., 461, 60

\bibitem[{Lovell {et~al.}(2017)}]{Lovell:2016nkp}
---. 2017, Mon. Not. Roy. Astron. Soc., 468, 2836

\bibitem[{Madau {et~al.}(1999)Madau, Haardt, \& Rees}]{Madau:1998cd}
Madau, P., Haardt, F., \& Rees, M.~J. 1999, Astrophys. J., 514, 648

\bibitem[{Menci {et~al.}(2016)Menci, Grazian, Castellano, \&
  Sanchez}]{Menci:2016eui}
Menci, N., Grazian, A., Castellano, M., \& Sanchez, N.~G. 2016, Astrophys. J.,
  825, L1

\bibitem[{Mesinger {et~al.}(2014)Mesinger, Ewall-Wice, \&
  Hewitt}]{Mesinger:2013nua}
Mesinger, A., Ewall-Wice, A., \& Hewitt, J. 2014, Mon. Not. Roy. Astron. Soc.,
  439, 3262

\bibitem[{{Miralda-Escud{\'e}} {et~al.}(2000){Miralda-Escud{\'e}}, {Haehnelt},
  \& {Rees}}]{Miralda:2000mhr}
{Miralda-Escud{\'e}}, J., {Haehnelt}, M., \& {Rees}, M.~J. 2000, \apj, 530, 1

\bibitem[{Moore {et~al.}(1999)}]{Moore:1999nt}
Moore, B. {et~al.} 1999, Astrophys. J., 524, L19

\bibitem[{Nakama {et~al.}(2017)Nakama, Chluba, \&
  Kamionkowski}]{Nakama:2017ohe}
Nakama, T., Chluba, J., \& Kamionkowski, M. 2017

\bibitem[{Okamoto {et~al.}(2008)Okamoto, Gao, \& Theuns}]{Okamoto:2008sn}
Okamoto, T., Gao, L., \& Theuns, T. 2008, Mon. Not. Roy. Astron. Soc., 390, 920

\bibitem[{Ouchi {et~al.}(2017)}]{Ouchi:2017cqe}
Ouchi, M. {et~al.} 2017

\bibitem[{Pacucci {et~al.}(2013)Pacucci, Mesinger, \& Haiman}]{Pacucci:2013jfa}
Pacucci, F., Mesinger, A., \& Haiman, Z. 2013, Mon. Not. Roy. Astron. Soc.,
  435, L53

\bibitem[{{Parsons} {et~al.}(2014){Parsons}, {Liu}, {Aguirre}, {Ali},
  {Bradley}, {Carilli}, {DeBoer}, {Dexter}, {Gugliucci}, {Jacobs}, {Klima},
  {MacMahon}, {Manley}, {Moore}, {Pober}, {Stefan}, \& {Walbrugh}}]{21cm:pla14}
{Parsons}, A.~R., {Liu}, A., {Aguirre}, J.~E., {Ali}, Z.~S., {Bradley}, R.~F.,
  {Carilli}, C.~L., {DeBoer}, D.~R., {Dexter}, M.~R., {Gugliucci}, N.~E.,
  {Jacobs}, D.~C., {Klima}, P., {MacMahon}, D.~H.~E., {Manley}, J.~R., {Moore},
  D.~F., {Pober}, J.~C., {Stefan}, I.~I., \& {Walbrugh}, W.~P. 2014, \apj, 788,
  106

\bibitem[{{Pober} {et~al.}(2015){Pober}, {Ali}, {Parsons}, {McQuinn},
  {Aguirre}, {Bernardi}, {Bradley}, {Carilli}, {Cheng}, {DeBoer}, {Dexter},
  {Furlanetto}, {Grobbelaar}, {Horrell}, {Jacobs}, {Klima}, {Kohn}, {Liu},
  {MacMahon}, {Maree}, {Mesinger}, {Moore}, {Razavi-Ghods}, {Stefan},
  {Walbrugh}, {Walker}, \& {Zheng}}]{21cm:pap15}
{Pober}, J.~C., {Ali}, Z.~S., {Parsons}, A.~R., {McQuinn}, M., {Aguirre},
  J.~E., {Bernardi}, G., {Bradley}, R.~F., {Carilli}, C.~L., {Cheng}, C.,
  {DeBoer}, D.~R., {Dexter}, M.~R., {Furlanetto}, S.~R., {Grobbelaar}, J.,
  {Horrell}, J., {Jacobs}, D.~C., {Klima}, P.~J., {Kohn}, S.~A., {Liu}, A.,
  {MacMahon}, D.~H.~E., {Maree}, M., {Mesinger}, A., {Moore}, D.~F.,
  {Razavi-Ghods}, N., {Stefan}, I.~I., {Walbrugh}, W.~P., {Walker}, A., \&
  {Zheng}, H. 2015, \apj, 809, 62

\bibitem[{Polisensky \& Ricotti(2014)}]{Polisensky:2013ppa}
Polisensky, E. \& Ricotti, M. 2014, Mon. Not. Roy. Astron. Soc., 437, 2922

\bibitem[{{Pritchard} \& {Loeb}(2012)}]{21cm:pl12}
{Pritchard}, J.~R. \& {Loeb}, A. 2012, Reports on Progress in Physics, 75,
  086901

\bibitem[{Robertson {et~al.}(2015)Robertson, Ellis, Furlanetto, \&
  Dunlop}]{Robertson:2015uda}
Robertson, B.~E., Ellis, R.~S., Furlanetto, S.~R., \& Dunlop, J.~S. 2015,
  Astrophys. J., 802, L19

\bibitem[{Rudd {et~al.}(2008)Rudd, Zentner, \& Kravtsov}]{Rudd:2007zx}
Rudd, D.~H., Zentner, A.~R., \& Kravtsov, A.~V. 2008, Astrophys. J., 672, 19

\bibitem[{Sawala {et~al.}(2013)Sawala, Frenk, Crain, Jenkins, Schaye, Theuns,
  \& Zavala}]{Sawala:2012cn}
Sawala, T., Frenk, C.~S., Crain, R.~A., Jenkins, A., Schaye, J., Theuns, T., \&
  Zavala, J. 2013, Mon. Not. Roy. Astron. Soc., 431, 1366

\bibitem[{Sawala {et~al.}(2016)}]{Sawala:2015cdf}
Sawala, T. {et~al.} 2016, Mon. Not. Roy. Astron. Soc., 457, 1931

\bibitem[{Schewtschenko {et~al.}(2016)Schewtschenko, Baugh, Wilkinson, Bœhm,
  Pascoli, \& Sawala}]{Schewtschenko:2015rno}
Schewtschenko, J.~A., Baugh, C.~M., Wilkinson, R.~J., Bœhm, C., Pascoli, S.,
  \& Sawala, T. 2016, Mon. Not. Roy. Astron. Soc., 461, 2282

\bibitem[{Schneider {et~al.}(2012)Schneider, Smith, Macci{\`o}, \&
  Moore}]{Schneider:2011yu}
Schneider, A., Smith, R.~E., Macci{\`o}, A.~V., \& Moore, B. 2012, Mon. Not.
  Roy. Astron. Soc., 424, 684

\bibitem[{Schneider {et~al.}(2013)Schneider, Smith, \&
  Reed}]{Schneider:2013ria}
Schneider, A., Smith, R.~E., \& Reed, D. 2013, Mon. Not. Roy. Astron. Soc.,
  433, 1573

\bibitem[{Schultz {et~al.}(2014)Schultz, Oñorbe, Abazajian, \&
  Bullock}]{Schultz:2014eia}
Schultz, C., Oñorbe, J., Abazajian, K.~N., \& Bullock, J.~S. 2014, Mon. Not.
  Roy. Astron. Soc., 442, 1597

\bibitem[{Seljak {et~al.}(2006)Seljak, Makarov, McDonald, \&
  Trac}]{Seljak:2006qw}
Seljak, U., Makarov, A., McDonald, P., \& Trac, H. 2006, Phys. Rev. Lett., 97,
  191303

\bibitem[{Somerville {et~al.}(2003)Somerville, Bullock, \&
  Livio}]{Somerville:2003sh}
Somerville, R.~S., Bullock, J.~S., \& Livio, M. 2003, Astrophys. J., 593, 616

\bibitem[{{Tinker} {et~al.}(2008){Tinker}, {Kravtsov}, {Klypin}, {Abazajian},
  {Warren}, {Yepes}, {Gottl{\"o}ber}, \& {Holz}}]{Tinker:2008tkk}
{Tinker}, J., {Kravtsov}, A.~V., {Klypin}, A., {Abazajian}, K., {Warren}, M.,
  {Yepes}, G., {Gottl{\"o}ber}, S., \& {Holz}, D.~E. 2008, \apj, 688, 709

\bibitem[{Viel {et~al.}(2013)Viel, Becker, Bolton, \& Haehnelt}]{Viel:2013apy}
Viel, M., Becker, G.~D., Bolton, J.~S., \& Haehnelt, M.~G. 2013, Phys. Rev.,
  D88, 043502

\bibitem[{Viel {et~al.}(2005)Viel, Lesgourgues, Haehnelt, Matarrese, \&
  Riotto}]{Viel:2005qj}
Viel, M., Lesgourgues, J., Haehnelt, M.~G., Matarrese, S., \& Riotto, A. 2005,
  Phys. Rev., D71, 063534

\bibitem[{Viel {et~al.}(2006)Viel, Lesgourgues, Haehnelt, Matarrese, \&
  Riotto}]{Viel:2006kd}
---. 2006, Phys. Rev. Lett., 97, 071301

\bibitem[{Viel {et~al.}(2008)}]{Viel:2007mv}
Viel, M. {et~al.} 2008, Phys. Rev. Lett., 100, 041304

\bibitem[{Vogelsberger {et~al.}(2012)Vogelsberger, Zavala, \&
  Loeb}]{Vogelsberger:2012ku}
Vogelsberger, M., Zavala, J., \& Loeb, A. 2012, Mon. Not. Roy. Astron. Soc.,
  423, 3740

\bibitem[{Wang \& White(2007)}]{Wang:2007he}
Wang, J. \& White, S. D.~M. 2007, Mon. Not. Roy. Astron. Soc., 380, 93

\bibitem[{Wang {et~al.}(2016)}]{Wang:2016rio}
Wang, L. {et~al.} 2016

\bibitem[{Y{\`e}che {et~al.}(2017)Y{\`e}che, Palanque-Delabrouille, Baur, \&
  BourBoux}]{Yeche:2017upn}
Y{\`e}che, C., Palanque-Delabrouille, N., Baur, J.~., \& BourBoux, H. d. M.~d.
  2017

\bibitem[{Yoshida {et~al.}(2003)Yoshida, Sokasian, Hernquist, \&
  Springel}]{Yoshida:2003rm}
Yoshida, N., Sokasian, A., Hernquist, L., \& Springel, V. 2003, Astrophys. J.,
  591, L1

\bibitem[{Yue \& Chen(2012)}]{Yue:2012na}
Yue, B. \& Chen, X. 2012, Astrophys. J., 747, 127

\end{thebibliography}

\end{document}